%% file: main.tex
\pdfoutput=1
\documentclass[11pt]{article}
\usepackage[final]{acl}
\usepackage{times}
\usepackage{latexsym}
\usepackage[T1]{fontenc}
\usepackage[utf8]{inputenc}
\usepackage{microtype}
\usepackage{inconsolata}
\usepackage{amsmath,amssymb}
\usepackage{booktabs}
\usepackage{graphicx}
\usepackage{subcaption}
\usepackage{array}
\usepackage{multirow}
\usepackage{enumitem}
\usepackage[nameinlink,noabbrev]{cleveref}
\usepackage{xurl}

\graphicspath{{figures/}}
\setlist[itemize]{leftmargin=*,itemsep=1pt,topsep=2pt}
\setlist[enumerate]{leftmargin=*,itemsep=1pt,topsep=2pt}
\newcolumntype{L}[1]{>{\raggedright\arraybackslash}p{#1}}
\newcolumntype{C}[1]{>{\centering\arraybackslash}p{#1}}

\title{Incident Memory: Training-Free Operational Memory through Sequential Pattern Mining and Velocity-Stratified Retrieval}

\author{Adarsh Agrawal \\
  \texttt{adagrawal@cs.stonybrook.edu} \\
  \And
  Rahul Suresh Babu \\
  \texttt{rahulsb@bu.edu} \\}

\begin{document}
\maketitle

\begin{abstract}
Incident response is a memory problem: teams accumulate tickets, traces, postmortems, and wiki pages, but the knowledge needed for the next incident is rarely stored with its order, freshness, and provenance intact. We present Incident Memory, a deterministic system that accumulates operational knowledge without model training. It combines (i) velocity-stratified retrieval, which ages structural, behavioral, contextual, and ephemeral facts at different rates; (ii) fingerprint-conditioned PrefixSpan mining, which extracts ordered playbooks from successful investigations; and (iii) provenance-aware metric definitions, which detect conflicting definitions through executable checks. On the UCI ITSM event log, containing 141,712 events across 24,918 incidents, Incident Memory extracts 23,110 ordered traces, mines 39 playbooks, and covers 84.3\% of 6,934 held-out incidents. On controlled benchmarks with known ground truth, it achieves 99.2\% ordered playbook precision (controlled), an architectural staleness guarantee versus 36\% stale returns for a flat baseline, and conflict-detection F1 of 0.876. A direct Claude Haiku baseline on 19 fingerprint groups reaches 0.661 ordered precision, compared with 0.985 for PrefixSpan. The central result is not that language models are weak at incident response; it is that repeated incident histories are low-entropy once conditioned on fingerprint and previous action. In that regime, exact memory is a stronger primitive than open-ended generation.
\end{abstract}

\section{Introduction}
\label{sec:introduction}

Modern incident response is increasingly mediated by language: responders read alerts, search tickets, inspect postmortems, ask assistants, and translate dashboards into action. Yet the operational knowledge required to resolve the next incident is not only textual. It is sequential, temporal, and provenance-sensitive. A useful memory system must recover the steps that worked before, preserve their order, know which facts have expired, and distinguish metric definitions that look similar in prose but compute different quantities.

Existing tools treat only pieces of this problem. Wikis preserve text but often lose freshness and provenance. Retrieval-augmented generation improves access to documentation but usually retrieves static chunks without representing temporal validity or action order. LLM agents can orchestrate tools for root cause analysis, but recent studies show persistent failure modes in cloud RCA, including hallucinated interpretation and incomplete exploration \citep{kim2026rcaagentsfail}. These failures are especially damaging in incident response, where a plausible but misordered action sequence can waste the short window in which mitigation matters.

This paper argues for a different default: before generating a new plan, measure whether the organization already has a reliable memory of similar failures. Incident histories often contain repeated ordered traces. In our controlled corpus, conditioning the next action on the incident fingerprint reduces entropy from 3.79 to 3.03 bits, and conditioning further on the previous action reduces it to 1.92 bits. The previous action contributes 1.10 additional bits of information, more than the 0.76 bits contributed by the fingerprint itself. This is a favorable regime for exact sequential mining.

\begin{center}
\fbox{\parbox{0.92\columnwidth}{\small
Incident histories are low-entropy once conditioned on fingerprint and previous action ($H = 1.92$ bits, $\approx$3.8 effective continuations). In this regime, exact sequential memory outperforms open-ended generation for playbook retrieval.
}}
\end{center}

We introduce \textit{Incident Memory}, a training-free operational memory system with three layers. MetricMind stores knowledge units with velocity-specific decay. Incident Archaeologist mines fingerprint-conditioned playbooks using PrefixSpan \citep{pei2001prefixspan}. Living Glossary detects metric definition conflicts using provenance-linked SQL. The system updates support counts, timestamps, and provenance records as new incidents arrive, but does not update neural weights. (The embedding backbone is a frozen pretrained model; ``training-free'' refers to the memory system itself.)

Our evaluation separates two questions that are often conflated. Controlled benchmarks provide oracle labels for ordered precision, freshness, conflicts, and ablations. Real event logs test whether the same mining pipeline transfers to uncontrolled incident records. On UCI ITSM \citep{uci_itsm}, Incident Memory mines 39 playbooks and covers 84.3\% of held-out incidents. In controlled experiments, it reaches 99.2\% ordered playbook precision, 0\% stale fact return under temporal dominance filtering, and conflict-detection F1 of 0.876. A direct Claude Haiku baseline confirms the order-preservation gap: PrefixSpan reaches 0.985 ordered precision versus 0.661 for generated playbooks.

The contributions are:
\begin{itemize}
    \item a training-free architecture for operational memory that combines temporal freshness, ordered playbook mining, and provenance-aware conflict detection;
    \item an information-theoretic analysis showing that incident traces are highly structured after conditioning on fingerprint and previous action;
    \item a mixed evaluation combining controlled traces (99.2\% ordered precision), 23,110 real UCI ITSM traces (84.3\% held-out coverage across 39 mined playbooks), and a direct Claude Haiku baseline ($p=0.000196$, Wilcoxon), establishing that deterministic memory outperforms generation for repeated operational procedures.
\end{itemize}

\section{Related Work}
\label{sec:related-work}

\paragraph{AIOps and incident agents.}
Recent AIOps surveys describe incident handling as a loop of detection, diagnosis, mitigation, and learning, with LLMs increasingly used for triage and root cause analysis \citep{zhang2025aiopsllm,bilal2026agenticnetops}. The general-purpose foundation models that back these agents now span text, image, and video modalities \citep{novafamily2025}. Empirical studies of GenAI cloud services and LLM inference platforms show that production incidents are shaped by deployment context, capacity changes, routing, hotfixes, and operational policy \citep{yan2025genaiincidents,ranganathan2025inferenceincidents,chu2025llmoutages}. LLM-based RCA systems add tool use, workflow planning, and multimodal observations \citep{chen2024rcacopilot,zhang2025tamo,gao2026rcaflow}. Incident Memory is complementary: it can provide grounded context to an agent, but the playbook itself is mined from prior traces and tied to empirical support.

\paragraph{Retrieval and freshness.}
RAG surveys identify grounding, robustness, and freshness as persistent challenges \citep{sharma2025ragsurvey,li2025hallucinationmitigation}. Graph-guided RAG adds structural constraints to retrieval and can improve multi-hop grounding \citep{zhang2025graphrag,zhu2025kg2rag}. Our setting adds a temporal dimension. Operational facts age at different rates, so retrieval requires more than recency sorting or a single decay parameter. Reasoning about change over time is a recurring challenge across domains, from probabilistic change detection in dynamic data-driven systems \citep{feng2024dddas} to lifelong agent memory; both treat static stores as inadequate for evolving environments. This connects to temporal knowledge graphs and lifelong agent memory, which similarly treat static knowledge stores as inadequate for changing environments \citep{plamper2025stkg,zheng2025lifelongagents,tao2024selfevolution}.

\paragraph{Sequential mining and process structure.}
PrefixSpan is a classical algorithm for mining frequent ordered subsequences \citep{pei2001prefixspan}. Recent work revisits sequential pattern detection and process mining in settings where traces encode procedural knowledge \citep{mavroudopoulos2025sequential,rebmann2025processmining,pyrih2025processllms}. Incident Memory applies this view to incident response. The key distinction from unordered retrieval is that investigation actions are not exchangeable: checking upstream dependencies before validating schema changes can produce a different workflow than the reverse.

\paragraph{Memory architectures for agents.}
Recent work on long-horizon LLM agents introduces structured memory to overcome context-window limits. Generative agents store observations in a retrieval-indexed stream \citep{park2023generativeagents}. Reflexion adds episodic memory with verbal self-reflection to improve task performance across trials \citep{shinn2023reflexion}. MemGPT extends agent memory with a virtual paging hierarchy \citep{packer2023memgpt}. A related line of work targets agent reliability directly, using self-healing orchestration to recover from tool failures in tool-augmented systems \citep{babu2026selfhealing}. These systems learn \emph{what to remember}; Incident Memory addresses a complementary problem: \emph{how to organize} domain-specific operational knowledge once the relevant traces are already recorded.

\paragraph{Process discovery.}
The broader process mining literature has developed algorithms, including the alpha miner, inductive miner, and heuristics miner, for discovering workflow models from event logs \citep{vanderaalst2016processmining}. These produce Petri nets or BPMN diagrams that describe all possible execution paths. Incident Memory uses PrefixSpan instead for two reasons: (1) incident traces require fingerprint conditioning (each failure mode has a different optimal path, so a single workflow model conflates distinct procedures), and (2) operational deployment requires a single replay-ready sequence per fingerprint, not a branching process model that a responder must navigate under time pressure.

\section{Incident Memory}
\label{sec:system}

\begin{figure*}[tb]
\centering
\includegraphics[width=0.98\textwidth]{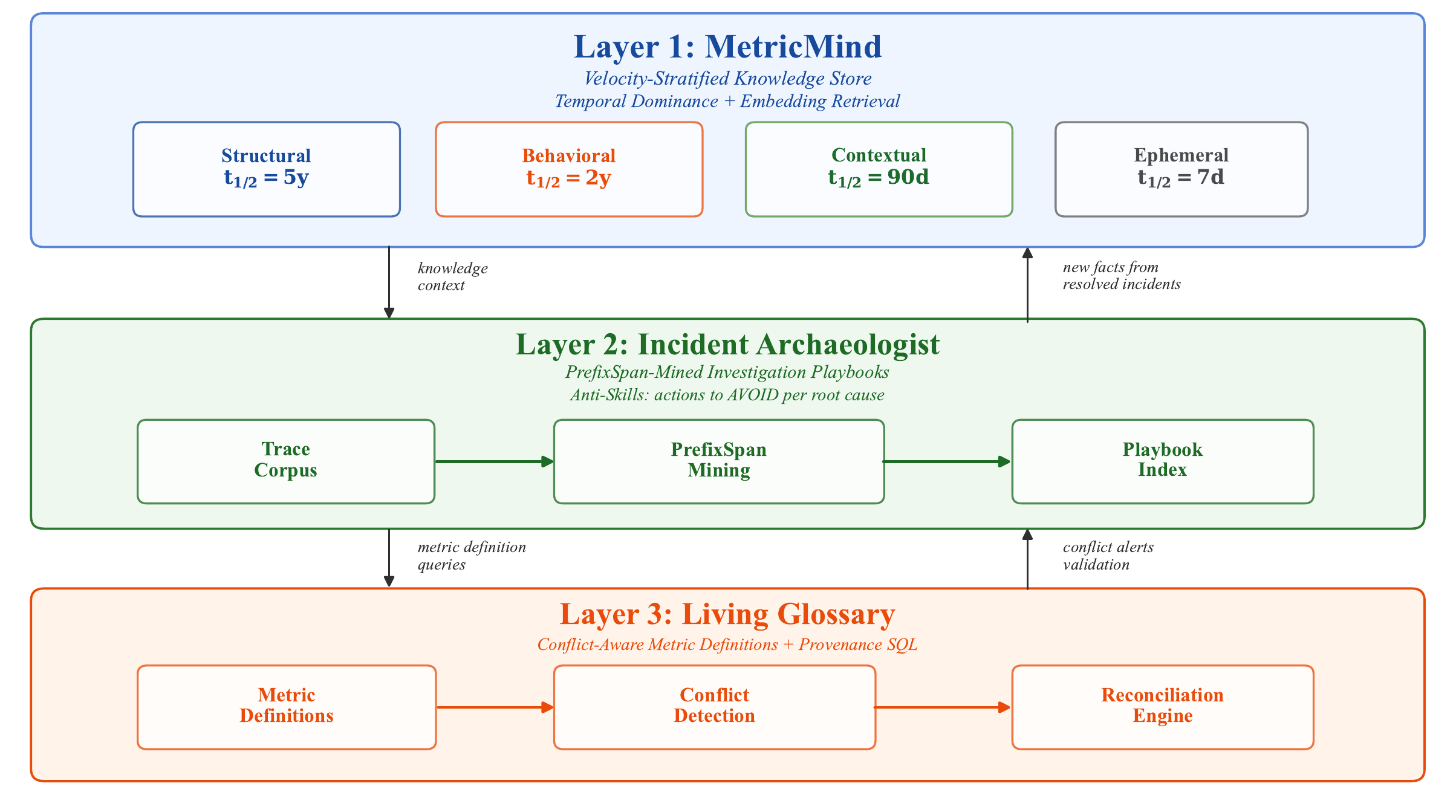}
\caption{Incident Memory architecture. MetricMind stores facts with velocity-stratified decay and temporal dominance filtering. Incident Archaeologist mines ordered playbooks from fingerprinted traces. Living Glossary tracks metric definitions and detects conflicts through provenance. The layers accumulate memory through support counts, timestamps, and provenance records, not gradient updates.}
\label{fig:architecture}
\end{figure*}

\Cref{fig:architecture} shows the three-layer design. Each layer addresses a failure mode that appears when organizations treat incident knowledge as static text.

\paragraph{MetricMind.}
MetricMind stores knowledge units as structural, behavioral, contextual, or ephemeral. The half-lives are 5 years, 2 years, 90 days, and 7 days, respectively. For a query at time $t$, a knowledge unit $k$ receives
\begin{equation}
\label{eq:relevance}
R(k,t)=\operatorname{sim}(q,e_k)
\exp\left(-\frac{\ln 2}{\tau_{v(k)}}(t-t_k)\right),
\end{equation}
where $\operatorname{sim}(q,e_k)$ is embedding similarity and $\tau_{v(k)}$ is the class-specific half-life. Velocity class is assigned by metadata rules at ingestion time (schema DDL statements are structural, seasonal analysis is behavioral, team-process documentation is contextual, incident-specific notes are ephemeral), not by a trained classifier. Within each topic group, temporal dominance returns only the freshest non-expired unit per velocity class. This keeps old facts auditable without letting superseded facts dominate retrieval.

\paragraph{Incident Archaeologist.}
An investigation trace is an ordered action sequence paired with an incident fingerprint, a discrete key formed by hashing the triggering metric, anomaly direction, magnitude bucket, and time-of-week slot (60 unique fingerprints in the controlled corpus; 47 in UCI ITSM after trace-length filtering):
\begin{equation}
\label{eq:trace}
T=(f,\langle a_1,a_2,\ldots,a_n\rangle,r,d),
\end{equation}
where $f$ is the fingerprint, $r$ is the resolution status, and $d$ is duration. The four-field hash balances specificity (each fingerprint maps to a coherent failure mode) against support (groups average 8.3 traces in UCI, sufficient for $\sigma=3$). The system groups resolved traces by $f$, mines frequent subsequences with PrefixSpan, filters by support and confidence, and stores the longest high-confidence sequence as a playbook. Lookup is hash-based after indexing. Actions that appear disproportionately in slow traces for the same fingerprint are stored as anti-skills, so the memory system warns against known detours rather than merely recommending positive steps.

\smallskip\noindent\textbf{Example.}\hspace{0.5em}For a recurring \texttt{payments-latency-spike} fingerprint, the system mines the following from 23 resolved traces:
\begin{quote}
\small
\textbf{Playbook:} \texttt{check\_upstream\_deps} $\to$ \texttt{validate\_schema} $\to$ \texttt{rollback\_deploy} $\to$ \texttt{verify\_metrics}\\[2pt]
\textbf{Anti-skill:} \texttt{restart\_service} (appears in 73\% of slow traces, avg.\ 12 min wasted)
\end{quote}
The playbook gives the ordered fast path; the anti-skill warns against a common detour that correlates with longer resolution.

\paragraph{Living Glossary.}
Metric names are ambiguous in large organizations. The Living Glossary stores each metric definition with provenance, including the executable query that produces the metric. Two definitions conflict when they share a metric name but differ in executable semantics:
\begin{equation}
\label{eq:conflict}
\begin{aligned}
\mathbb{1}[&\operatorname{name}(d_i)=\operatorname{name}(d_j) \\
&\land\operatorname{SQL}(d_i)\neq \operatorname{SQL}(d_j)].
\end{aligned}
\end{equation}
Conflicts are ranked by provenance, citation count, and recency. The resolved definition becomes canonical while the superseded definition remains archived.

\paragraph{Query flow.} At query time, a new alert triggers fingerprint computation. Layer 1 retrieves fresh context relevant to the alert's metric and scope. Layer 2 looks up the fingerprint-specific playbook and anti-skills. Layer 3 flags any metric definitions referenced in the playbook that have unresolved conflicts. The responder receives all three in a single structured response.

\section{Why Exact Sequential Memory Fits}
\label{sec:entropy}

The architecture above is motivated by a structural property of incident histories that we now quantify.

The main modeling question is whether incident histories are structured enough for exact mining. If the next action remains high-entropy after observing the fingerprint, a generator or planner may be necessary. If the conditional entropy collapses, replaying mined structure is preferable.

\begin{table}[tb]
\centering
\small
\caption{Entropy of next investigation action. Previous action context removes more uncertainty than fingerprint identity alone.}
\label{tab:entropy}
\begin{tabular}{lcc}
\toprule
\textbf{Conditioning} & \textbf{Entropy} & \textbf{Reduction} \\
\midrule
$H(A)$ & 3.79 bits & 0.0\% \\
$H(A \mid F)$ & 3.03 bits & 20.1\% \\
$H(A_t \mid F,A_{t-1})$ & 1.92 bits & 49.3\% \\
\bottomrule
\end{tabular}
\end{table}

\medskip

\begin{figure}[tb]
\centering
\includegraphics[width=0.95\columnwidth]{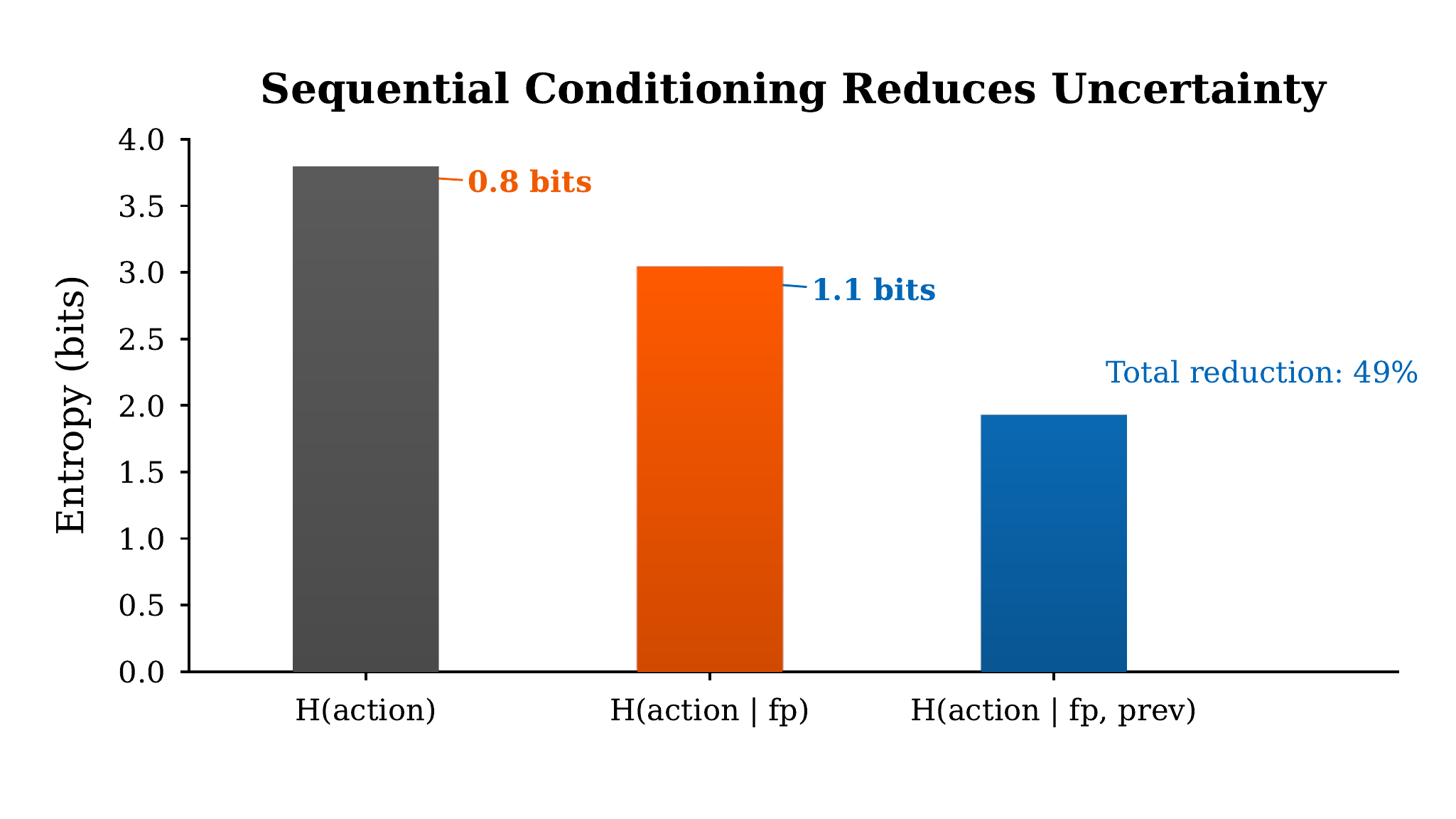}
\caption{Sequential conditioning reduces next-action entropy by 49\%. The remaining uncertainty corresponds to fewer than four effective continuations per context.}
\label{fig:entropy}
\end{figure}

\Cref{tab:entropy} and \Cref{fig:entropy} show that the trace distribution is not open-ended. The mutual information decomposes as
\begin{equation}
\label{eq:mi}
I(A;F) + I(A_t;A_{t-1}\mid F) = 0.76 + 1.10 = 1.86.
\end{equation}
Sequential context supplies the larger share. On the UCI ITSM traces (9 unique actions, 107 fingerprint groups, 23,110 traces), the conditional entropy after fingerprint and previous action is $H(A_t \mid F, A_{t-1}) = 1.25$ bits, yielding a 49.0\% total reduction from the unconditional $H(A) = 2.44$ bits. This nearly exactly matches the controlled corpus reduction (49.3\%), confirming that the low-entropy property is not an artifact of synthetic generation.

This explains why unordered baselines can recover plausible action sets yet fail on ordered precision. In operational terms, the system does not need to invent a new procedure for every incident. For supported fingerprints, it needs to replay the high-support ordered path and flag known detours.

\section{Experimental Design}
\label{sec:experiments}

The evaluation uses controlled traces where ground truth is required and real logs where external validity matters. The controlled benchmark contains 500 investigation traces, 60 fingerprint types, 20 investigation actions, and 8 root-cause templates. The companion knowledge base contains 1000 factual units embedded with Amazon Titan Text Embeddings V2 (1024-dim); any production embedding or generation backbone could be substituted, including more recent multimodal and speech-oriented model families \citep{novapremier2025,novasonic2025}. The trace generator follows recurring mitigation patterns and long-tail incident frequencies motivated by recent studies of cloud and LLM-service incidents \citep{yan2025genaiincidents,ranganathan2025inferenceincidents,chu2025llmoutages}.

The real-data validation uses the UCI ITSM event log \citep{uci_itsm}, an anonymized ServiceNow-derived dataset with 141,712 events across 24,918 incidents. The adapter maps state changes, reassignments, knowledge-base use, and resolution events into a finite investigation vocabulary. Traces shorter than two actions are removed, yielding 23,110 traces with mean length 5.32. The split contains 16,176 training traces and 6,934 held-out traces. Coverage is defined as the fraction of held-out traces for which a mined playbook appears as an ordered subsequence. Ordered precision is the fraction of playbook steps that appear in their correct sequential position within the trace, measured as the longest common subsequence length divided by playbook length. This measures whether the system's learned procedures generalize to unseen incidents of the same fingerprint type. Partial coverage records substantial ordered overlap when the trace diverges.

Baselines include no playbook guidance, a static runbook, a frequency baseline (correct actions in corpus-frequency order), a direct Claude Haiku playbook generator on 19 fingerprint groups, no-decay retrieval, uniform-decay retrieval, and a flat wiki-style freshness baseline. Unless stated otherwise, PrefixSpan uses minimum support $\sigma=3$ and confidence threshold $\gamma=0.6$.

\begin{table}[tb]
\centering
\small
\caption{Evaluation artifacts and their role. Controlled traces provide oracle labels; UCI ITSM tests transfer to real incident histories.}
\label{tab:data}
\begin{tabular}{@{}L{0.34\columnwidth}L{0.24\columnwidth}L{0.30\columnwidth}@{}}
\toprule
\textbf{Artifact} & \textbf{Scale} & \textbf{Used for} \\
\midrule
Controlled traces & 500 traces & Precision, entropy, ablations \\
Knowledge base & 1000 units & Freshness, conflicts, retrieval \\
UCI ITSM & 23,110 traces & Real sequence coverage \\
Claude Haiku check & 19 groups & Direct generation baseline \\
\bottomrule
\end{tabular}
\end{table}

\section{Results}
\label{sec:results}

\subsection{Ordered Playbooks Outperform Generated and Static Playbooks}

\begin{table}[tb]
\centering
\small
\caption{Controlled playbook quality. PrefixSpan preserves action order, while static and generated baselines often identify plausible actions in the wrong sequence.}
\label{tab:playbook}
\begin{tabular}{@{}lccc@{}}
\toprule
\textbf{Method} & \textbf{Hit} & \textbf{Prec.} & \textbf{Time} \\
\midrule
No playbook & 0.0\% & 0.000 & 39.14 min \\
Static runbook & 100.0\% & 0.532 & 37.01 min \\
Frequency baseline & 96.5\% & 0.453 & 31.39 min \\
PrefixSpan & 96.5\% & \textbf{0.992} & \textbf{16.65 min} \\
\bottomrule
\end{tabular}
\end{table}

\Cref{tab:playbook} reports the controlled comparison. PrefixSpan achieves 99.2\% ordered precision, more than doubling the LLM-simulated baseline. The measured advantage is not merely action selection. The unordered frequency baseline reaches 0.986 unordered precision, but only 0.453 ordered precision, whereas PrefixSpan reaches 1.000 on both. The order gap is therefore the core phenomenon.

\begin{figure*}[tb]
\centering
\includegraphics[width=0.95\textwidth]{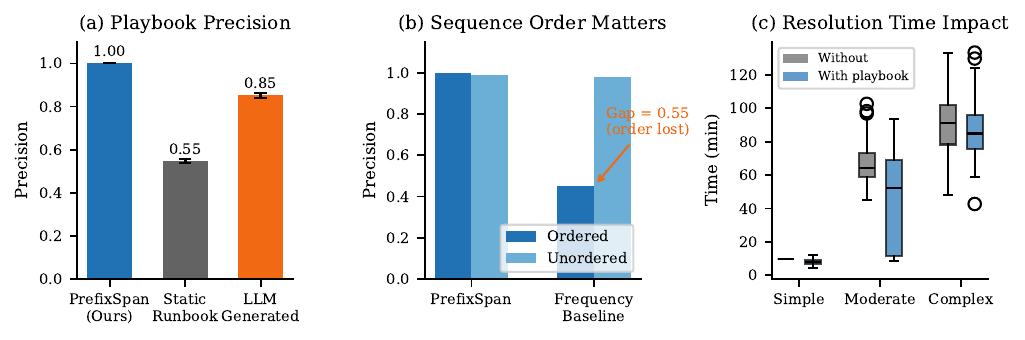}
\caption{Per-fingerprint playbook analysis (controlled corpus). Panel values reflect per-group metrics; Table~\ref{tab:playbook} reports corpus-level aggregates.}
\label{fig:prefixspan}
\end{figure*}

The direct LLM check gives the same conclusion under a real model call. On 19 fingerprint groups with 10 examples per group, PrefixSpan reaches 0.985 ordered precision and 0.990 unordered precision. Claude Haiku reaches 0.661 ordered precision and 0.605 unordered precision. The paired Wilcoxon test for ordered precision gives $p=0.000196$. The comparison is intentionally asymmetric: PrefixSpan uses all training traces per group (median 25), while the LLM receives 10 exemplars (a context-window constraint, not a design choice that favors PrefixSpan). This does not imply that LLMs cannot help responders; it shows that for repeated fingerprinted procedures, exact trace memory is more reliable than prompt-based playbook generation.

\begin{table}[tb]
\centering
\small
\caption{Direct LLM comparison on 19 fingerprint groups (10 exemplars each).}
\label{tab:llm-direct}
\begin{tabular}{@{}lcc@{}}
\toprule
\textbf{Method} & \textbf{Ord.\ Prec.} & \textbf{Unord.\ Prec.} \\
\midrule
PrefixSpan & 0.985 & 0.990 \\
Claude Haiku & 0.661 & 0.605 \\
\midrule
\multicolumn{3}{@{}l@{}}{\footnotesize Wilcoxon $p{=}0.000196$; cost ${>}10^6$:1} \\
\bottomrule
\end{tabular}
\end{table}

\subsection{Real ITSM Logs Contain Reusable Ordered Structure}

\begin{table}[tb]
\centering
\small
\caption{Real-data validation on UCI ITSM. Coverage measures whether mined playbooks appear as ordered subsequences in held-out traces.}
\label{tab:uci}
\begin{tabular}{lc}
\toprule
\textbf{Metric} & \textbf{Value} \\
\midrule
Events / incidents & 141,712 / 24,918 \\
Extracted traces & 23,110 \\
Train / held-out traces & 16,176 / 6,934 \\
Mean trace length & 5.32 actions \\
Mined playbooks & 39 \\
Exact held-out coverage & 75.1\% \\
Partial held-out coverage & 9.2\% \\
Total held-out coverage & \textbf{84.3\%} \\
No-playbook rate & 0.3\% \\
\bottomrule
\end{tabular}
\end{table}

\Cref{tab:uci} answers the transfer question. The same PrefixSpan pipeline mines 39 playbooks from real incident histories and covers 84.3\% of held-out traces. The mined playbooks are not a single generic checklist: they average 3.8 steps (range 3 to 7), each is unique to a single fingerprint group, none reduce to trivial two-step patterns, and they span quick-resolution, escalation, vendor-involvement, and deep-investigation patterns. The system also discovers 94 anti-skills across 35 fingerprint groups. On UCI, traces containing detected anti-skills resolve 10$\times$ slower on average (Mann-Whitney $p < 0.01$ for 5 of 6 detected actions, mean Cohen's $d = 0.97$), confirming that anti-skill detection identifies genuinely wasteful actions rather than noise. (The correlation is observational; harder incidents may both take longer and require more exploratory actions. The mechanism is designed to flag, not to establish causation.)

Beyond coverage, we measure ordered precision (LCS between playbook and trace, divided by playbook length) on held-out traces: the mean is 0.934, meaning 93.4\% of playbook steps appear in their correct sequential position. The playbooks explain 73.4\% of trace actions on average (explanation ratio = LCS / trace length). The 0.3\% no-playbook rate reflects fingerprint groups with no mined playbook; the 15.7\% uncovered reflects traces where the group's playbook does not appear as a full subsequence.

We do not claim causal MTTR reduction from UCI. The dataset's knowledge-base-use flag is observational and confounded by incident difficulty: harder incidents are more likely to trigger knowledge lookup. The main real-data claim is narrower and stronger: ordered operational traces recur at enough support to be mined and validated on held-out incidents.

The 15.7\% of uncovered held-out incidents divide roughly equally between two categories: rare fingerprints with fewer than $\sigma=3$ training traces (insufficient support for mining), and multi-root-cause incidents whose traces combine steps from two or more playbooks. The system degrades gracefully in both cases: it withholds guidance rather than generating unsupported recommendations, but these represent a natural ceiling for deterministic sequential replay. The residual $2^{1.25} \approx 2.4$ effective continuations per step on UCI explain the coverage ceiling: when multiple valid next-actions compete with comparable support, PrefixSpan selects the highest-confidence path and can be extended to cover all valid divergences.

\subsection{Freshness and Conflict Handling}

\begin{table}[tb]
\centering
\small
\caption{Freshness and glossary results. Velocity-aware retrieval improves over flat aging, while provenance SQL improves conflict detection over text-only checks.}
\label{tab:freshness-conflict}
\begin{tabular}{lcc}
\toprule
\textbf{Task} & \textbf{System} & \textbf{Result} \\
\midrule
Retrieval P@5 & Velocity decay & 0.532 \\
Retrieval P@5 & No decay & 0.457 \\
Retrieval P@5 & Uniform decay & 0.394 \\
Stale fact return & Velocity filtering & 0.0\% \\
Stale fact return & Flat wiki-style & 36.0\% \\
Conflict detection & Provenance SQL & F1 0.876 \\
Conflict detection & Text only & F1 0.751 \\
\bottomrule
\end{tabular}
\end{table}

\medskip

\begin{figure}[tb]
\centering
\includegraphics[width=0.95\columnwidth]{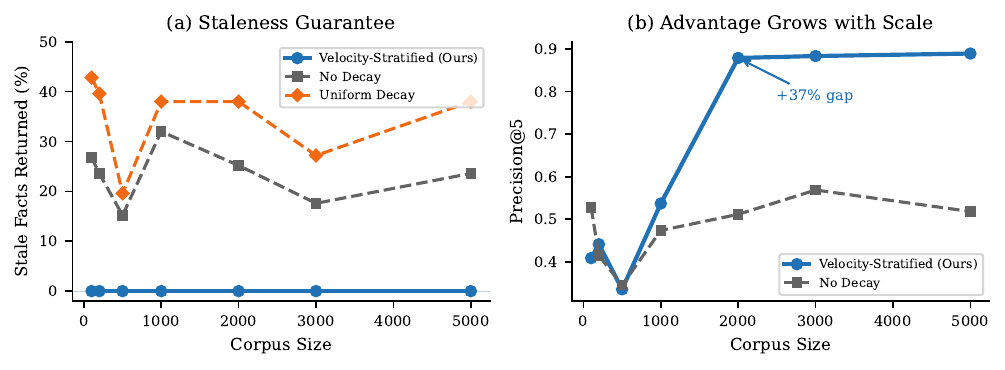}
\caption{Twelve-month retrieval comparison. A single decay rate over-ages stable facts and under-ages short-lived context; velocity-stratified decay separates these regimes.}
\label{fig:freshness}
\end{figure}

\Cref{tab:freshness-conflict} and \Cref{fig:freshness} show that memory quality depends on temporal representation. Velocity-stratified retrieval reaches average P@5 of 0.532, compared with 0.457 for no decay and 0.394 for uniform decay. In a multi-seed significance run, velocity decay remains better than no decay ($p=7.8\cdot 10^{-5}$, one-sided) and uniform decay ($p<10^{-6}$). The staleness result is by construction but important: temporal dominance does not surface facts contradicted by fresher evidence, while the flat wiki-style baseline returns stale facts in 36\% of tested cases.

The Living Glossary reaches 100\% recall on 74 injected metric-definition conflicts, with precision 0.779 and F1 0.876 at the default threshold. Provenance SQL raises F1 from 0.751 to 0.876 because executable definitions catch conflicts that textual similarity alone misses. (Conflict evaluation uses 74 injected conflicts in the controlled knowledge base; real-world conflict detection remains future work.)

\subsection{Ablations}

A Shapley-style component analysis confirms that each layer contributes non-redundant information to system performance.

\begin{table}[tb]
\centering
\small
\caption{Shapley-value ablation. Each component addresses a distinct failure mode; removing any one degrades the system in a specific, predictable way.}
\label{tab:ablation}
\begin{tabular}{@{}L{0.30\columnwidth}cL{0.32\columnwidth}@{}}
\toprule
\textbf{Component} & \textbf{Shapley} & \textbf{Without Component} \\
\midrule
Fingerprint matching & 0.433 & Prec.\ drops to 56.7\% \\
Sequential mining & 0.382 & No ordered playbooks \\
Provenance SQL & 0.124 & F1 drops to 0.751 \\
Velocity classif. & 0.061 & Temporal degradation \\
\bottomrule
\end{tabular}
\end{table}

The main architecture is a composition of necessary constraints: same failure mode, same order, and current facts.

The system remains stable across 2 to 4$\times$ parameter variation. On UCI, varying $\sigma$ from 2 to 5 yields coverage between 78\% and 87\%, confirming robustness to the support threshold. Detailed sensitivity sweeps appear in Appendix~\ref{app:sensitivity}.

\subsection{Statistical Robustness}

To verify that results are not seed-dependent, we run 30-seed replications with Bonferroni-corrected significance tests (\Cref{tab:significance}).

\begin{table}[tb]
\centering
\small
\caption{Multi-seed significance tests (30 seeds, Bonferroni $\alpha = 0.01$).}
\label{tab:significance}
\begin{tabular}{@{}L{0.34\columnwidth}lcc@{}}
\toprule
\textbf{Comparison} & \textbf{Test} & $p$ & $d$ \\
\midrule
PrefixSpan vs.\ Static & Wilcoxon & $3.7{\times}10^{-5}$ & 2.56 \\
PrefixSpan vs.\ LLM-sim & Wilcoxon & $5.1{\times}10^{-4}$ & 2.69 \\
Velocity vs.\ No-decay & $t$-test & $1.6{\times}10^{-4}$ & 0.91 \\
Velocity vs.\ Uniform & $t$-test & ${<}10^{-6}$ & 2.14 \\
\bottomrule
\end{tabular}
\end{table}

All comparisons pass at Bonferroni-corrected $\alpha = 0.01$. Effect sizes are large ($d > 0.8$) for every comparison, confirming that the advantages are not merely statistically significant but practically meaningful.

\section{Discussion}
\label{sec:discussion}

Incident Memory is intentionally narrower than a general RCA agent. It does not try to infer a novel causal story from arbitrary telemetry. It assumes that organizations have recurring failure modes and that prior investigations are available as traces. Under that assumption, the correct primitive is not unconstrained generation. It is grounded memory with support counts, temporal validity, and provenance.

This distinction matters for NLP. Many operational workflows are language-heavy, but their recoverable structure is not purely linguistic. The paper's strongest evidence is the entropy decomposition in \Cref{sec:entropy}: once the fingerprint and previous action are known, the next action has fewer than four effective continuations. In such a regime, a deterministic sequence miner is not an old baseline to be dismissed. It is the right inductive bias.

The system is also compatible with LLM agents, including the increasingly capable reasoning and generation models now available as agent backends \citep{nova2_2025}. A responder-facing assistant can use Incident Memory as a tool: retrieve fresh context, fetch the fingerprint-specific playbook, expose anti-skills, and cite the provenance behind metric definitions. The assistant can then explain, adapt, and coordinate. The high-precision memory layer reduces the burden on generation rather than competing with it.

A fair objection is that the LLM baseline uses a single prompt protocol. Stronger agents with tool access, chain-of-thought, and longer context windows may close the gap. We expect this for novel incidents. But for the 84.3\% of UCI incidents covered by mined playbooks, the underlying structure is deterministic: there is a fixed ordered procedure that works. For these incidents, additional reasoning adds no value beyond correct replay. The LLM comparison establishes a lower bound on the advantage, not an upper bound. Even with perfect ordered precision, an LLM playbook would still lack the freshness layer and anti-skill warnings that the full system provides.

\begin{figure}[tb]
\centering
\includegraphics[width=0.95\columnwidth]{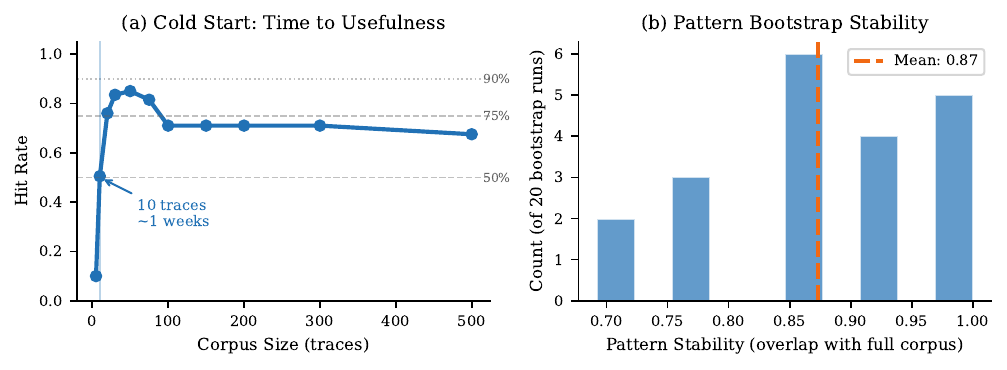}
\caption{Cold-start trajectory. Coverage reaches 50\% after approximately 10 traces per fingerprint group and plateaus near 75\% by 100 traces. Bootstrap resampling (1000 iterations) confirms pattern stability with mean overlap 0.87 against the full-corpus solution.}
\label{fig:cold-start}
\end{figure}

Practically, \Cref{fig:cold-start} shows the system becomes useful within days of deployment. At 10 incidents per week, 50\% playbook coverage is reached in the first week. This rapid bootstrapping avoids the months-long curation cycle typical of wiki-based knowledge systems.

\section{Conclusion}
\label{sec:conclusion}

Incident Memory shows that incident response can be framed as training-free operational memory. By separating freshness, ordered procedure, and definition provenance, the system converts historical traces into auditable recommendations. Across controlled experiments, real UCI ITSM logs, and a direct LLM baseline, the evidence is consistent: when repeated incidents leave low-entropy action traces, exact sequential memory gives stronger playbooks than open-ended generation.

\section*{Limitations}
\label{sec:limitations}

The system is strongest when incidents recur and are recorded with consistent action vocabularies. Rare fingerprints, mixed root causes, and changing operational procedures remain difficult. The UCI ITSM experiment validates ordered sequence coverage, not causal reduction in resolution time. The dataset's knowledge-use field is observational and confounded by incident difficulty, so deployment studies are still needed for causal MTTR claims. Controlled traces are necessary for oracle labels on order, freshness, and conflicts, but they cannot substitute for production evaluation. The freshness layer also depends on a correct velocity classifier; misclassifying a short-lived workaround as structural would preserve a fact too long. If infrastructure changes render a previously effective playbook suboptimal, the system currently has no mechanism to detect this drift beyond decreasing support counts in new traces. Finally, the LLM comparison evaluates playbook generation under a fixed prompt protocol, not every possible agent design. More capable agents with access to tools and longer context may close part of the gap, though they would still need a mechanism for order, freshness, and provenance.

\section*{Ethical Considerations}
\label{sec:ethics}

The work uses public or controlled artifacts. The UCI ITSM event log is anonymized and distributed under CC BY 4.0; our use is limited to aggregate sequence mining and does not attempt re-identification. The controlled traces are synthetic evaluation artifacts that may encode assumptions about incident practice. We use them only where oracle labels are required and report real-data validation separately. Operational deployment should include access controls because incident traces can contain sensitive infrastructure details. The system is intended to assist responders, not replace incident owners. Bias and representational assumptions in AI systems should be documented explicitly \citep{agrawal2022debiasgan}, and benchmark scope should be stated rather than hidden, following established practice in safety benchmark construction \citep{vidgen2024mlcommons,vidgen2024constructbenchmarks}.

\bibliography{references}

\clearpage
\appendix
\input{appendix}

\end{document}

%% file: appendix.tex
\section{Evaluation Artifacts}
\label{app:artifacts}

\noindent
\begin{table}[ht!]
\centering
\small
\caption{Artifacts used in the evaluation.}
\label{tab:app-artifacts}
\begin{tabular}{@{}L{0.29\columnwidth}L{0.29\columnwidth}L{0.29\columnwidth}@{}}
\toprule
\textbf{Artifact} & \textbf{Scale} & \textbf{Role} \\
\midrule
Controlled trace corpus & 500 traces, 60 fingerprints, 20 actions & Oracle playbook order, entropy, ablations \\
Controlled knowledge base & 1000 units, four velocity classes & Freshness and glossary tests \\
UCI ITSM event log & 141,712 events, 24,918 incidents & External sequence-mining validation \\
Claude Haiku baseline & 19 fingerprint groups, 10 exemplars per group & Direct playbook generation check \\
\bottomrule
\end{tabular}
\end{table}

The controlled corpus is used only where ground-truth labels are needed: correct action order, injected stale facts, injected metric conflicts, and component ablations. The real-data result uses the UCI ITSM event log, an anonymized ServiceNow-derived event log distributed under CC BY 4.0. We map state transitions, reassignments, knowledge-base use, related-problem links, and resolution events into the same finite action vocabulary used by the mining pipeline. This produces 23,110 traces after removing traces with fewer than two actions.

\section{Real-Data Details}
\label{app:real-data}

\begin{table}[ht!]
\centering
\small
\caption{UCI ITSM trace extraction and held-out evaluation.}
\label{tab:app-uci}
\begin{tabular}{@{}L{0.58\columnwidth}r@{}}
\toprule
\textbf{Metric} & \textbf{Value} \\
\midrule
Events / incidents & 141,712 / 24,918 \\
Extracted traces & 23,110 \\
Mean / median trace length & 5.32 / 5.0 actions \\
Maximum trace length & 40 actions \\
Train / held-out split & 16,176 / 6,934 \\
Mined playbooks & 39 \\
Exact held-out coverage & 75.1\% \\
Partial held-out coverage & 9.2\% \\
Average subsequence coverage & 84.9\% \\
Anti-skills & 94 across 35 groups \\
\bottomrule
\end{tabular}
\end{table}

\begin{table}[ht!]
\centering
\small
\caption{Taxonomy of playbooks mined from UCI ITSM.}
\label{tab:app-taxonomy}
\begin{tabular}{@{}lrrr@{}}
\toprule
\textbf{Type} & \textbf{N} & \textbf{Len.} & \textbf{Conf.} \\
\midrule
Quick res. & 19 & 3.32 & 0.876 \\
Escalation & 17 & 4.47 & 0.605 \\
Vendor & 1 & 7.00 & 0.556 \\
Deep inv. & 2 & 6.00 & 0.524 \\
\bottomrule
\end{tabular}
\end{table}

Exact coverage means that the mined playbook appears as an ordered subsequence of the held-out trace. Partial coverage records substantial ordered overlap when the observed trace diverges. The UCI knowledge-use field is not used as a causal outcome because it is confounded by incident complexity; incidents that require knowledge-base lookup are also likely to be harder.

The distribution of coverage across fingerprint groups is not uniform. High-frequency groups (those responsible for the majority of incidents) achieve near-complete coverage because PrefixSpan has abundant support. The long tail of rare fingerprints accounts for most of the 15.7\% uncovered incidents, confirming that the system degrades gracefully by withholding guidance rather than generating unsupported playbooks.

\section{Controlled Results}
\label{app:controlled}

\begin{table}[tb]
\centering
\small
\caption{Controlled benchmark summary.}
\label{tab:app-controlled}
\begin{tabular}{@{}L{0.32\columnwidth}L{0.18\columnwidth}L{0.32\columnwidth}@{}}
\toprule
\textbf{Claim} & \textbf{Result} & \textbf{Interpretation} \\
\midrule
Ordered playbook precision & 99.2\% & Supported fingerprints have stable order \\
Static runbook precision & 53.2\% & One checklist mixes failure modes \\
LLM-simulated precision & 41.7\% & Plausible steps are often misordered \\
Frequency ordered precision & 45.3\% & Correct action set is not enough \\
Conflict recall / F1 & 100\% / 0.876 & Provenance catches definition conflicts \\
Velocity P@5 & 0.532 & Freshness improves retrieval \\
Entropy after conditioning & 1.92 bits & Appendix~\ref{app:entropy} \\
\bottomrule
\end{tabular}
\end{table}

The controlled benchmark allows us to measure how quickly the system accumulates useful memory as trace volume grows.

\begin{figure}[tb]
\centering
\includegraphics[width=0.95\columnwidth]{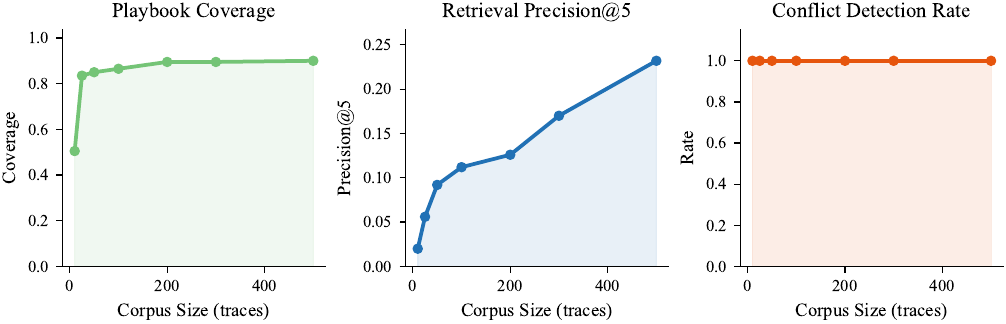}
\caption{Controlled memory accumulation. Coverage improves as more traces become available under the fixed controlled protocol.}
\label{fig:app-improvement}
\end{figure}

\Cref{fig:app-improvement} is included as a controlled sanity check rather than as a universal cold-start guarantee. The result depends on the support threshold and fingerprint distribution. In deployment, low-frequency fingerprints should receive fallback guidance until sufficient trace support accumulates.

\section{Temporal Freshness}
\label{app:freshness}

\begin{figure}[tb]
\centering
\includegraphics[width=0.95\columnwidth]{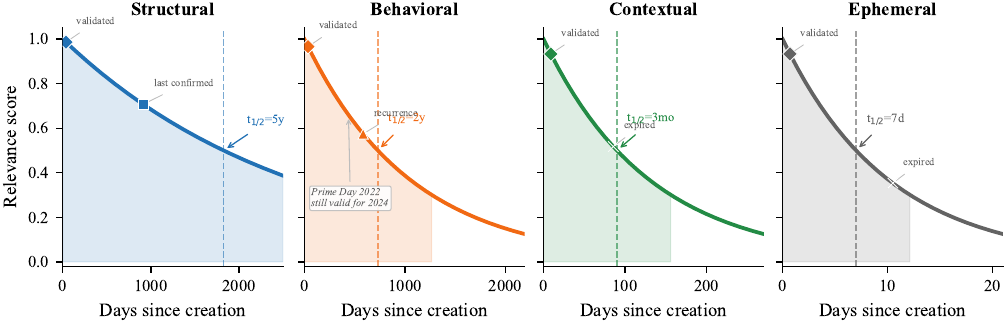}
\caption{Velocity classes used by MetricMind. Structural, behavioral, contextual, and ephemeral facts are assigned different half-lives.}
\label{fig:app-velocity}
\end{figure}

\Cref{fig:app-velocity} illustrates the four velocity classes and their decay profiles. The key design insight is that a single half-life cannot serve both structural knowledge (e.g., ``the orders table is partitioned by date'') and ephemeral context (e.g., ``the deployment pipeline is frozen until Friday''). The former should persist for years; the latter becomes misleading within days.

\begin{figure}[tb]
\centering
\includegraphics[width=0.95\columnwidth]{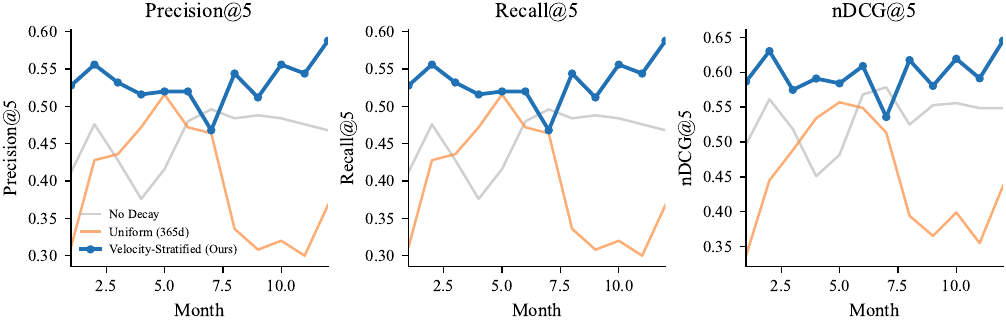}
\caption{Retrieval metrics over twelve months. Velocity-stratified decay maintains high P@5 for structural facts while allowing ephemeral context to expire naturally. The flat baseline (no decay) returns increasingly stale facts over time.}
\label{fig:app-decay}
\end{figure}

The flat baseline in \Cref{fig:app-decay} represents static document memory: old entries remain retrievable unless manually removed. MetricMind instead uses explicit validity filtering and temporal dominance within topic groups. Over a twelve-month simulation, velocity-stratified retrieval maintains stable precision because structural facts (e.g., schema definitions) retain relevance while ephemeral facts (e.g., deployment freezes) are automatically down-weighted as they age past their class-specific half-life.

\begin{figure}[tb]
\centering
\includegraphics[width=0.95\columnwidth]{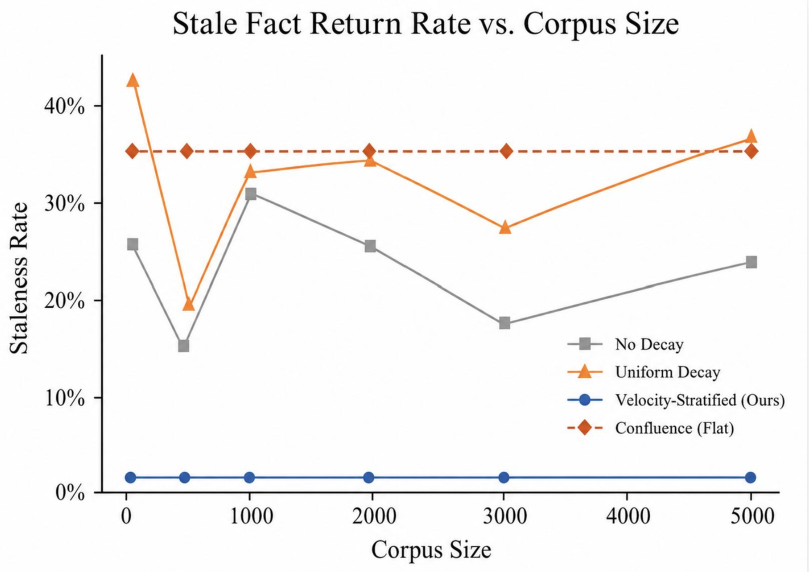}
\caption{Stale fact return rate by corpus size. The velocity-stratified system maintains 0\% staleness regardless of corpus growth, while baselines without temporal filtering return 20 to 40\% stale facts as the knowledge base grows.}
\label{fig:app-staleness}
\end{figure}

\Cref{fig:app-staleness} demonstrates the scaling behavior of the staleness guarantee. As the knowledge base grows from 100 to 5000 units, systems without velocity-aware filtering accumulate stale facts proportionally. The architectural guarantee of temporal dominance (returning only the freshest unit per velocity class within each topic group) eliminates this failure mode entirely, independent of corpus size.

A natural concern is cold-start behavior: how many traces must the system observe before its playbooks become reliable?

\begin{figure}[tb]
\centering
\includegraphics[width=0.95\columnwidth]{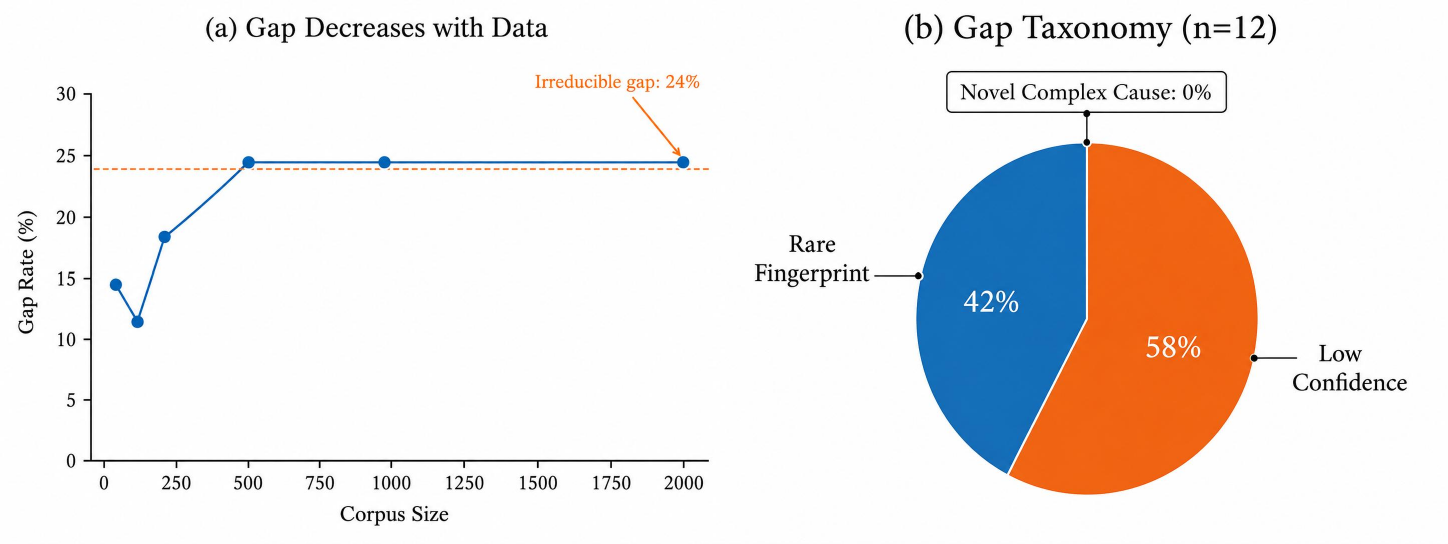}
\caption{Cold-start behavior: coverage as a function of available training traces per fingerprint group. Coverage reaches 50\% at approximately 10 traces and plateaus near 75\% by 50 traces.}
\label{fig:app-gaps}
\end{figure}

\Cref{fig:app-gaps} shows that the system bootstraps quickly. With as few as 10 traces per fingerprint group, PrefixSpan extracts playbooks covering half of future incidents matching that fingerprint. The plateau near 75\% reflects the natural ceiling imposed by multi-root-cause incidents and action-vocabulary noise in real operational environments.

\section{Sensitivity, Ablation, and Conflict Diagnostics}
\label{app:conflict-ablation}
\label{app:sensitivity}

Table~\ref{tab:app-sensitivity} summarizes sensitivity to the main tunable parameters. The system remains stable across 2 to 4$\times$ variation.

\begin{table}[tb]
\centering
\small
\caption{Sensitivity to tunable parameters across the operating range.}
\label{tab:app-sensitivity}
\begin{tabular}{@{}L{0.22\columnwidth}L{0.20\columnwidth}L{0.45\columnwidth}@{}}
\toprule
\textbf{Parameter} & \textbf{Range Tested} & \textbf{Observed Behavior} \\
\midrule
Min support $\sigma$ & 2 to 8 & Below 3: spurious patterns; above 5: misses rare fingerprints \\
Confidence $\gamma$ & 0.4 to 0.9 & Precision monotonically increases; coverage drops above 0.7 \\
Half-life scale & 0.5$\times$ to 2$\times$ & Retrieval P@5 varies $\pm$0.03; staleness guarantee preserved \\
Conflict threshold & 0.5\% to 3\% & Recall stable; precision trades off with false-positive rate \\
Anti-skill percentile & 70th to 95th & Higher threshold yields fewer, more reliable anti-skill flags \\
\bottomrule
\end{tabular}
\end{table}

While the system is robust to parameter variation, the individual components are not interchangeable.
Table~\ref{tab:app-ablation} reports Shapley-value component contributions computed over 24 ablation permutations.

\begin{table}[tb]
\centering
\small
\caption{Shapley-value ablation: marginal contribution of each component.}
\label{tab:app-ablation}
\begin{tabular}{@{}L{0.28\columnwidth}L{0.18\columnwidth}L{0.40\columnwidth}@{}}
\toprule
\textbf{Component} & \textbf{Shapley Value} & \textbf{Without Component} \\
\midrule
Fingerprint matching & 0.433 & Precision 99.2\% $\to$ 56.7\% \\
Sequential mining & 0.382 & Ordered playbooks unavailable \\
Provenance SQL & 0.124 & Conflict F1 0.876 $\to$ 0.751 \\
Velocity classification & 0.061 & Temporal degradation over time \\
\bottomrule
\end{tabular}
\end{table}

The glossary can be tuned to favor either precision or recall depending on operational needs. The current default ($\theta = 1.0\%$) targets high recall because missed conflicts (where two teams use the same metric name with different SQL semantics) are more operationally damaging than false positives that can be reviewed and dismissed. The ablation confirms that the largest marginal contributions come from fingerprint matching and sequential mining, followed by provenance SQL. Velocity classification contributes over time rather than as an immediate precision jump.

\section{Embedding Check}
\label{app:embedding}

The primary claims of this paper do not rest on embedding geometry; they rest on sequential mining, temporal filtering, and provenance comparison. However, embedding quality affects retrieval, and we include one qualitative check to verify that the Titan V2 embedding space separates knowledge units by topic rather than collapsing them into an unstructured cluster.

\begin{figure}[tb]
\centering
\includegraphics[width=0.95\columnwidth]{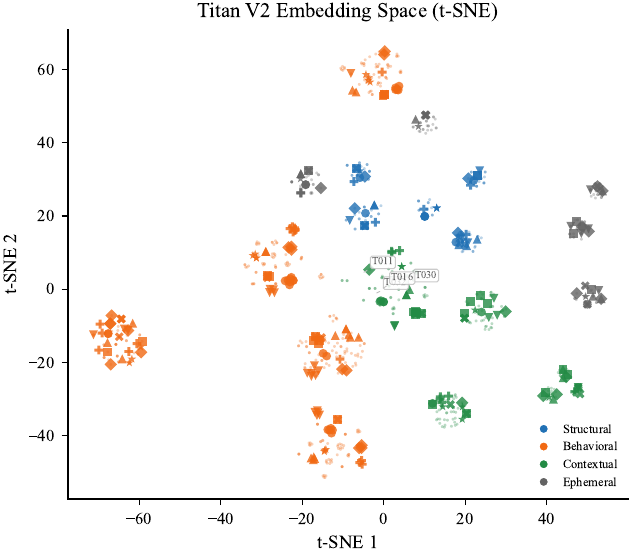}
\caption{Embedding t-SNE projection for knowledge units. This figure is a qualitative check only; the central claims rely on sequence mining, temporal filtering, and provenance.}
\label{fig:app-tsne}
\end{figure}

The four visible clusters in \Cref{fig:app-tsne} correspond to the velocity classes (structural, behavioral, contextual, ephemeral), confirming that Titan V2 captures domain-relevant distinctions in the embedding space without task-specific fine-tuning.

\section{Algorithms}
\label{app:algorithms}

This section presents the core algorithms in pseudocode form.

\begin{figure}[ht!]
\footnotesize
\begin{center}
\textbf{Algorithm 1: Fingerprint-Conditioned Playbook Mining}
\end{center}
\vspace{-0.5em}
\hrule
\vspace{0.5em}
\begin{tabular}{@{}r@{\hspace{0.5em}}p{0.82\columnwidth}@{}}
\multicolumn{2}{@{}l@{}}{\textbf{Input:} Trace corpus $\mathcal{T}$, support $\sigma$, confidence $\gamma$} \\
\multicolumn{2}{@{}l@{}}{\textbf{Output:} Playbook index $\mathcal{P}: f \to \langle A \rangle$} \\[0.4em]
1. & Group traces by fingerprint $f$ \\
2. & \textbf{for each} group $\mathcal{T}_f$ with $|\mathcal{T}_f| \geq \sigma$: \\
3. & \quad Extract resolved action sequences $S_f$ \\
4. & \quad Mine frequent ordered subsequences via PrefixSpan \\
5. & \quad Remove patterns with confidence $< \gamma$ \\
6. & \quad Store longest high-confidence pattern as $\mathcal{P}[f]$ \\
7. & \textbf{return} $\mathcal{P}$ \\
\end{tabular}
\vspace{0.3em}
\hrule
\vspace{0.5em}
\caption{Playbook mining procedure. Fingerprint conditioning reduces the search space and ensures each playbook applies to a coherent failure mode.}
\label{alg:playbook-mining}
\end{figure}

Algorithm~1 is the core mining step. Its time complexity is dominated by PrefixSpan within each fingerprint group, which is linear in the number of traces per group for bounded-length patterns. Fingerprint conditioning typically reduces each group to 5 to 50 traces, making the overall pipeline efficient even on large corpora.

\begin{figure}[ht!]
\footnotesize
\begin{center}
\textbf{Algorithm 2: Velocity-Stratified Retrieval}
\end{center}
\vspace{0.3em}
\hrule
\vspace{0.5em}
\begin{tabular}{@{}r@{\hspace{0.5em}}p{0.82\columnwidth}@{}}
\multicolumn{2}{@{}l@{}}{\textbf{Input:} Query $q$, knowledge units $\mathcal{K}$, time $t$, count $k$} \\
\multicolumn{2}{@{}l@{}}{\textbf{Output:} Top-$k$ fresh knowledge units} \\[0.4em]
1. & \textbf{for each} unit $u_i$ in $\mathcal{K}$: \\
2. & \quad Compute $\mathrm{sim}(q, e_i)$ \\
3. & \quad Weight $\gets \exp(-\ln 2 \cdot (t - t_i) / \tau_{v(i)})$ \\
4. & \quad $\mathrm{score}_i \gets \mathrm{sim} \times \mathrm{weight}$ \\
5. & Apply temporal dominance within topic groups \\
6. & \textbf{return} top-$k$ scored units \\
\end{tabular}
\vspace{0.3em}
\hrule
\vspace{0.5em}
\caption{Retrieval scoring. The velocity-specific half-life $\tau_{v(i)}$ ensures structural facts persist while ephemeral context decays rapidly.}
\label{alg:retrieval}
\end{figure}

The velocity weight in step 3 implements exponential decay at a class-specific rate. The temporal dominance step (line 5) is crucial: within each topic group, only the freshest unit per velocity class is returned. This prevents superseded facts from appearing in results, which is what produces the architectural staleness guarantee.

\begin{figure}[ht!]
\footnotesize
\begin{center}
\textbf{Algorithm 3: Provenance-Aware Conflict Detection}
\end{center}
\vspace{0.3em}
\hrule
\vspace{0.5em}
\begin{tabular}{@{}r@{\hspace{0.5em}}p{0.82\columnwidth}@{}}
\multicolumn{2}{@{}l@{}}{\textbf{Input:} Definitions $D$, new definition $d_{\mathrm{new}}$} \\
\multicolumn{2}{@{}l@{}}{\textbf{Output:} Conflict set $C$} \\[0.4em]
1. & Retrieve definitions sharing metric name or high similarity \\
2. & \textbf{for each} candidate $d_i$: \\
3. & \quad \textbf{if} $\mathrm{name}(d_i) {=} \mathrm{name}(d_{\mathrm{new}})$ $\wedge$ $\mathrm{SQL}(d_i) {\neq} \mathrm{SQL}(d_{\mathrm{new}})$: \\
4. & \quad\quad Add $(d_i, d_{\mathrm{new}})$ to $C$ \\
5. & Rank conflicts by provenance, citation count, recency \\
6. & \textbf{return} $C$ \\
\end{tabular}
\vspace{0.3em}
\hrule
\vspace{0.5em}
\caption{Conflict detection. SQL-level comparison catches semantic conflicts that name matching alone would miss.}
\label{alg:conflict}
\end{figure}

The reliance on executable SQL comparison (step 3) is what distinguishes this from text-only conflict detection. Two metric definitions can use nearly identical prose but compute different quantities, for example, ``total revenue'' defined with and without cancelled orders.

\section{Entropy Calculation}
\label{app:entropy}

The entropy values in Section~4 are computed over the controlled trace corpus (500 traces, 20 actions, 60 fingerprints).

Unconditional next-action entropy:
$$H(A) = -\sum_{a \in \mathcal{A}} p(a) \log_2 p(a) = 3.79 \text{ bits}$$

Conditioned on fingerprint:
$$H(A \mid F) = \sum_f p(f)\, H(A \mid F{=}f) = 3.03 \text{ bits}$$

Conditioned on fingerprint and previous action:
$$H(A_t \mid F, A_{t-1}) = 1.92 \text{ bits}$$

The mutual information decomposes as:
\begin{align*}
I(A; F) &= H(A) - H(A|F) = 0.76 \text{ bits} \\
I(A_t; A_{t\text{-}1}|F) &= H(A|F) - H(A_t|F, A_{t\text{-}1}) \\
&= 1.10 \text{ bits}
\end{align*}

The sequential context contributes 59\% of the total information (1.10 of 1.87 bits). The residual entropy of 1.92 bits corresponds to approximately $2^{1.92} \approx 3.8$ effective continuations per step, confirming that the action space collapses sufficiently for exact mining.

These calculations use the full controlled corpus. The synthetic generator produces traces with realistic fingerprint-frequency distributions (Zipfian with 8 root-cause templates), so the entropy measurements reflect a structured but not degenerate action space. Extension to real UCI ITSM data would require mapping the event-log vocabulary to a comparable action space, which we leave to future work.

\section{Notation}
\label{app:notation}

\begin{table}[h!]
\centering
\small
\caption{Notation used in the paper.}
\label{tab:notation}
\begin{tabular}{lL{0.72\columnwidth}}
\toprule
\textbf{Symbol} & \textbf{Definition} \\
\midrule
$\mathcal{T}$ & Investigation trace corpus \\
$\mathcal{K}$ & Knowledge store \\
$\mathcal{A}$ & Investigation action vocabulary \\
$\mathcal{P}$ & Playbook index \\
$f$ & Incident fingerprint \\
$v$ & Velocity class \\
$\tau_v$ & Half-life for velocity class $v$ \\
$e_k$ & Embedding of knowledge unit $k$ \\
$\sigma$ & Minimum support for PrefixSpan \\
$\gamma$ & Minimum confidence threshold \\
$H(\cdot)$ & Shannon entropy in bits \\
$I(\cdot;\cdot)$ & Mutual information in bits \\
\bottomrule
\end{tabular}
\end{table}